# MICROLENSING, or GALACTIC TWINKLING

N.W. Evans

*Theoretical Physics, Department of Physics, 1 Keble Road, Oxford OX1 3NP, UK*

Microlensing has established itself as a powerful new method for the detection of baryonic dark matter in the Galaxy. The theory of microlensing is sketched and its similarity with the optical effect of twinkling is explained. The bulk of the article presents a new analysis of the data-set on microlensing towards the Large Magellanic Cloud. The extent, flattening and velocity anisotropy of the Galactic halo are unknown. So, it is vital to analyse the microlensing data-set with families of models that span the viable ranges of these structural parameters. Also crucial is proper modelling of the dark matter halo of the Large Magellanic Cloud. Despite all the unknowns, a robust conclusion is that the Galactic and LMC haloes cannot be primarily built from objects in the mass range $10^{-7}\,{\rm M}_\odot$ - $0.1\,{\rm M}_\odot$. If the baryonic component of the Galactic halo is a fraction $f$ of the total mass, then $f \sim 0.4 - 0.5$ and the mass of the deflectors probably lies between $0.05\,{\rm M}_\odot - 1.0\,{\rm M}_\odot$. Stronger claims concerning the masses of the dark objects are unwarranted because the estimates are sensitive to the uncertain velocity anisotropy.

## 1 Introduction

Why do the stars twinkle? The stars twinkle because turbulence and mixing in the upper atmosphere causes patches of high and low refractive index. In a region of high refractive index, the wavefront loses; in a region of low refractive index, the wavefront gains. This changes a plane-parallel wavefront into a corrugated one. Where the wavefront is convex, the rays converge; where the wavefront is concave, the rays diverge. This concentrates the energy of the wave in certain patches.[1] Microlensing is galactic twinkling. Stellar images in the Large and Small Magellanic Clouds (LMC, SMC) and Galactic Bulge are microlensed because unseen masses in the Galaxy cause variations in the refractive index. A chance aligment of a dark object near the axis between an observer and source causes passing light rays to be deflected. Light is received by the observer from two paths. The observer records two images. In microlensing, the two images are separated by micro-arcseconds and cannot be resolved. Instead, the observer sees a time-dependent amplification of the image that is symmetric and achromatic.[2,3,4]

An early suggestion that microlensing offers a way to confirm the existence of baryonic objects in galaxy haloes was made by Petrou[5] while she was a graduate student at Cambridge University. This work forms a substantial chapter in her Ph. D. thesis, but was never published elsewhere. Amongst other things, Petrou explicitly notes that the stars in the Large Magellanic

Cloud are possible sources and correctly estimates the probability that any one star is microlensed is of the order $10^{-7}$. Discouraged by this low number, her thesis chapter ends poignantly with the words that "all these [phenomena] are quite unlikely ever to be observed, but it may be the only way to detect these invisible objects, if, after all, there are any around us."

Paczyński[6] has aptly described his daring 1986 paper[2] as a "science fiction proposal". Although microlensing of cosmological sources had been studied for over a decade, Paczyński noticed the advantages in bringing the sources nearer to home, namely that the timescales of the events come down. The down-side is that the probability any one source is microlensed is very low. Paczyński realised – and convinced the astronomical community – that the the monitoring of the light-curves of millions of stellar images in the Large Magellanic Cloud might now be technologically feasible. More or less directly inspired by Paczyński's audacity, the microlensing searches began.

## 2   The Microlensing Observables

An experimental group monitors a patch of the sky. At some galactic longitude and latitude, the group observes a rate of microlensing (a number of events per million stars monitored per year) and a distribution of timescales (two events with timescales between 10 and 12 days, four events between 12 and 14 days, and so on). The observables depend on two unknown distributions – the present day mass function (PDMF) of the lenses and the distribution of proper motions of the lens and source. At first sight, it seems we can say nothing definite – because both these distributions are poorly known. However, there is a neat detail that was first spotted by Press and Gunn[7] who studied microlensing in a cosmological context back in 1973. From these primary observables, we can construct a secondary observable – the optical depth to microlensing – which does not depend on the uncertain distributions, but only depends on the density of deflectors or lenses. First, a threshold amplification – say 1.34 – is picked above which a microlensing signal can be detected. For this choice, the impact parameter must be within an Einstein radius $R_{\rm E}$. If a lens comes within a distance $R_{\rm E}$ of the observer-source axis, a microlensing event is recorded. So, let us construct a microlensing tube, whose radius at any spot is $R_{\rm E}$.[3] The optical depth to microlensing is the number of lenses in the tube

$$\tau = \frac{1}{M} \int_0^{D_{\rm s}} \pi R_{\rm E}^2 \rho\, dx \qquad (1)$$

Here, the deflectors have density $\rho$ and characteristic mass $M$, while $D_{\rm s}$ is the distance from observer to source. The optical depth is independent of the distribution of velocities by construction. It is independent of the PDMF because $R_{\rm E}$ is proportional to the square root of the mass of the lens. It only depends on the density of deflectors. It has a natural interpretation as the probability that a given star is microlensed and so can be constructed directly from the observables:

$$\tau = \frac{\pi}{2} \sum_i \frac{t_i}{NT\epsilon_i} \qquad (2)$$

Here, N is the number of stars monitored for a period of time $T$, $t_i$ is the timescale of the $i$th event and $\epsilon_i$ is the efficiency of the experiment at that timescale. Part of the power of microlensing as a tool for studying Galactic structure comes from the robustness of the optical depth.

Which dark objects can cause microlensing? Microlensing searches can detect a variety of baryonic objects. These include: (1) stellar remnants, such as neutron stars, white dwarfs, red dwarfs and dim stars, (2) brown dwarfs, generally defined as stellar objects below the hydrogen-burning limit of $\sim 0.08\,{\rm M}_\odot$, (3) Jupiters, with a mass typically $\sim 0.001\,{\rm M}_\odot$, and (4) snowballs, which are have roughly the same mass as Jupiters, but are bound by molecular rather than gravitational forces. The present-day results of the microlensing experiments already set strong constraints on how much of the Galactic halo can be built from all these objects. There are two further baryonic objects which have been suggested as possible residents of haloes. The first of these – black holes or dark clusters of mass $\sim 10^4\,{\rm M}_\odot$ or greater [8] – produce events with timescales of the orders of decades and the present experiments offer no information on their existence. The second of these – clouds of cold molecular hydrogen [9] – may be too diffuse to cause events.

The original suggestion of Paczyński was to monitor the Large Magellanic Clouds – a nearby irregular galaxy at a distance $\sim 50\,{\rm kpc}$. It is viewed through the screen of dark objects in the halo of our own Galaxy. Two collaborations, the Macho and Eros groups, [10,11] reported the discovery of microlensing of LMC stars in 1993. A second target is the Galactic Bulge or bar. [12,13] Here, we look through the disk and outer Bulge of our own Galaxy towards windows unobscured by dust near the Galactic Center. Possible lenses include not just dark matter in the disk (if it exists), but also stars in the disk and Bulge. At present there are (at least) three groups monitoring fields in the Galactic Center and the total number of events observed in this direction exceeds a hundred. [14,15,16,17]

## 3  Analysis of the Magellanic Cloud Microlensing Dataset

The Macho group has published the results of 2.1 years of the photometry of 8.5 million stars.[18] They found 8 candidate events. The optical depth towards the Clouds is estimated to be $\sim 2.9 \times 10^{-7}$. The timescales of the events lie between one and one hundred days.

Let us analyse this dataset with the model introduced by Evans.[19] The sources lie in the disk of the LMC. The lenses may lie either in the disk and halo of the Milky Way or in the disk, halo and bar of the LMC.[20,21] A crude facsimile of the LMC is provided by embedding an inclined disk in a spherical dark halo to reproduce the observed tilt of the LMC disk ($\sim 45°$) and the observed position angle of its line of nodes, as well as its observed asymptotic rotation curve of amplitude $\sim 80\,\mathrm{kms^{-1}}$. The disk of the Milky Way is exponential with a scale-length of $\sim 3.5\,\mathrm{kpc}$ and an axis ratio of 1:12. It is instructive to consider two sets of models: canonical ones in which the local column density near the Sun is taken as $\sim 71\,\mathrm{M_\odot pc^{-2}}$ and maximal ones in which the local column density is $\sim 100\,\mathrm{M_\odot pc^{-2}}$. The former are in agreement with analyses of the vertical kinematics of tracer populations[22], the latter suggested by the microlensing data towards the Galactic Center.[14] The disk stars are taken as cold and rotating with the circular speed. The PDMF of the nearby disk population is reasonably well known.[23] Of course, the most important source of deflectors is the dark halo of our own Galaxy. Almost everything about the dark halo is unknown, including its ellipticity, its extent and its velocity anisotropy. The data must be analysed within the context of a family of models, each of which is a plausible representation for the Milky Way. Such a family of haloes is provided by *the power-law models*.[24] These are a class of simple and elegant solutions – with accompanying velocity distributions – to the self-gravitation equations discovered a couple of years ago. They have become the standard model of the Galactic halo for analyses of the microlensing data.[25,26,27]

For each model, the contribution to the differential rate with respect to timescale from deflectors in the LMC halo, disk and bar and the Galactic halo and disk is summed. This is multiplied by the efficiency of the experiment at that timescale, then integrated over all timescales to obtain the number of events that should have been observed. In fact, Alcock et al.[18] saw 8 events. Models predicting in excess of 14.5 events are excluded at the 95 per cent confidence level. Fig. 1 shows these excluded models if the disk is canonical. Plotted horizontally is the logarithm (to the base ten) of the mass of the deflectors in units of the solar mass. Plotted vertically are the unknowns. The velocity anisotropy of the halo could be radial (in these units, unity) or tangential. The flattening could vary between E0 (spherical) and E7 (highly flattened). The halo could extend to 30 kpc or it could extend to beyond the Magellanic clouds. For each model, the cross-hatched region shows the range of forbidden masses of the dark objects. The characteristic mass of the dominant component of the Galactic halo cannot lie within the forbidden regions. How do things change if the galactic disk is maximal? Fig. 2 shows that the excluded ranges shrink somewhat, By looking at the regions forbidden in all six panels of Figs 1 and 2, we deduce that – irrespective of the uncertainties – the characteristic mass of the major contributor to the Galactic halo cannot lie in the range $0.1\,M_\odot - 10^{-4}\,M_\odot$. But, there is still more information! The most stringent limits on short timescale events are provided by the Eros group.[28] They carried out 10 months of photometry of 82,00 stars with up to 46 measurements per night. They found no candidates for very short timescale events. This important null result provides very stringent limits on low mass deflectors. Again, models predicting in excess of 3.0 events can be excluded at the 95 percent confidence level. Looking at the data with the same ensemble of models, we find the excluded range runs from $10^{-4}\,M_\odot - 10^{-7}\,M_\odot$ irrespective of the uncertainties of the modelling (see Fig. 3). Combining both the Eros and the Macho data tells us that a broad swathe of masses from $0.1\,M_\odot - 10^{-7}\,M_\odot$ are excluded as the dominant constituent of Galactic haloes.

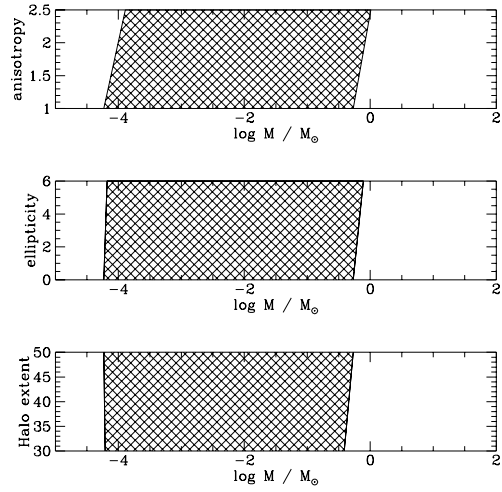

Figure 1: The horizontal axis plots the logarithm of the mass of the deflectors in units of the solar mass, while the vertical axes show the uncertainties – the anisotropy, the halo extent and the flattening. The characteristic mass of the dominant component of Galactic haloes does not lie in the cross-hatched regions.

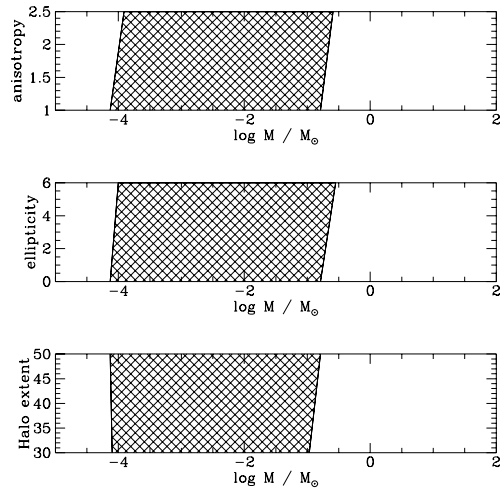

Figure 2: As Fig. 1, but for a maximal disk model.

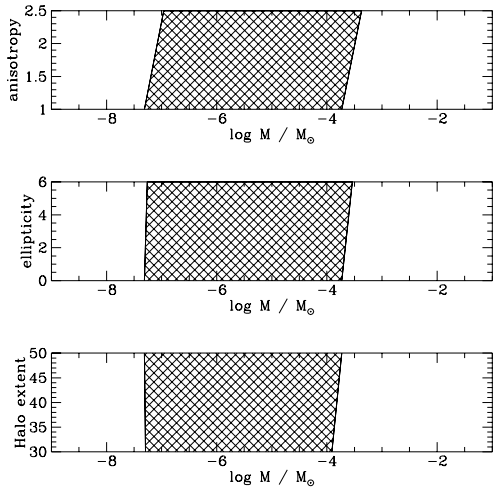

Figure 3: As Fig. 1, but for the data-set of Aubourg et al.

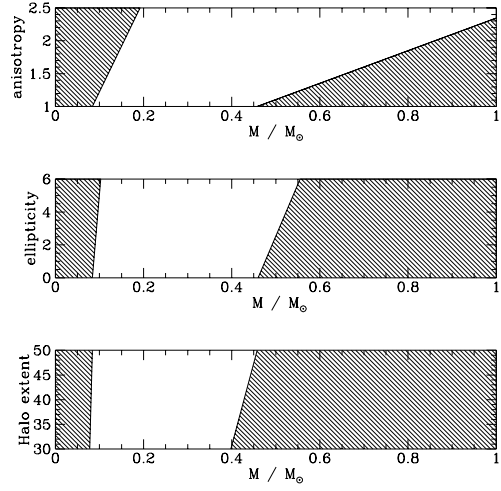

Figure 4: Models that reproduce both the rate and the optical depth are shown as the unshaded windows. The horizontal axis shows the allowed masses of the deflectors, the vertical axis the uncertainties. The disk is canonical.

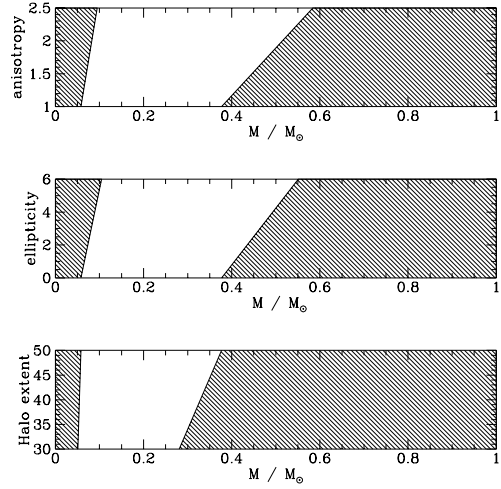

Figure 5: As Fig. 4, but for a maximal disk model.

Thus far, the halo has been assumed to be largely baryonic and only the information on the rate has been exploited. Perhaps more likely is that a certain fraction of the halo $f$ is baryonic. This baryonic population has a characteristic mass $M$. Suppose we now require that the models also reproduce the observed optical depth of $\tau \sim 2.9^{+1.4}_{-0.9} \times 10^{-7}$. What is the possible mass range of the deflectors as the uncertainties are varied? The answer is provided in Figs 4 and 5, which show the allowed mass ranges for canonical and maximal Galactic disks respectively. Again, the modelling uncertainties are plotted vertically and the mass of the lenses horizontally. Models in the unshaded windows can reproduce the rate and the optical depth to within the experimental uncertainties. For the purpose of inferring the masses of the deflectors, the most important uncertainty is clearly seen to be the anisotropy of the velocity distribution. This has a substantial effect on the inferred mass of the lens. If the velocity distribution is made more radial, then the average duration of microlensing events becomes larger as the lenses remain within the microlensing tube for longer. If the data is mistakenly analysed with an almost isotropic model, then the mass of the lenses is over-estimated. Note, too, that if the Galactic disk is maximal (as in Fig. 5), then the inferred mass of the lenses shrinks. This happens because the potential deep in the halo is reduced if the Galactic disk is maximal. The characteristic velocities of the dark lenses is reduced, and so the observed timescales can be reproduced with less massive lenses.

Bearing in mind all the modelling uncertainties, it is reasonable to make the following conclusion. If the Galactic disk is canonical, the baryon mass fraction $f$ is $\sim 0.4$ and the mass of the lenses is typically $.1\,{\rm M}_\odot - 1\,{\rm M}_\odot$. If the disk is maximal, the baryon mass fraction is $\sim 0.5$ and the mass of the lenses is $0.05\,{\rm M}_\odot - 0.6\,{\rm M}_\odot$. This range includes brown dwarfs as well as low mass stars and compact remnants. These mass ranges are somewhat lower than those deduced by other investigators[18] because the effects of the anisotropy of the velocity distribution have been considered.

## 4  Analysis of the Galactic Bulge Microlensing Dataset

Microlensing towards the Magellanic Clouds has the most direct relevance for the physics of dark matter, as the deflectors are unambiguously dark objects in the haloes of the Galaxy and the LMC. The microlensing searches towards the Bulge need not be directly probing Galactic dark matter. This is because the low mass end of the stellar luminosity function provides a known supply of deflectors. These are the dim stars either in the disk[12] or the Bulge[29]. The lenses could also be disk dark matter (if it exists) or halo dark matter (though the contribution of the latter towards the Galactic Center is surely small). The sources may lie in the Bulge or they may even lie behind the Bulge.[30]

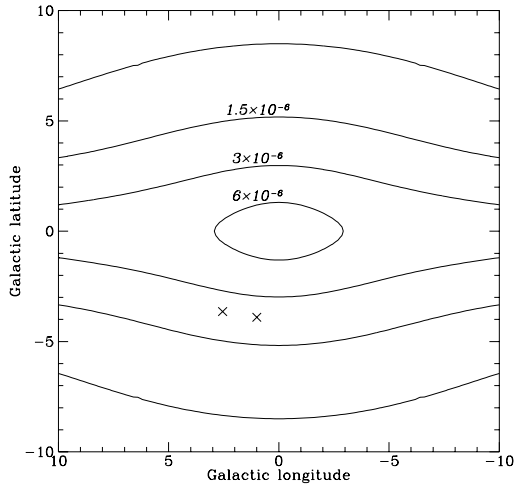

Figure 6: Contours of equal optical depth in the inner Galaxy, assuming an axisymmetric model with a heavy bulge and spheroid of mass $\sim 3.9 \times 10^{10} M_\odot$ and a disk of mass $\sim 5.6 \times 10^{10} M_\odot$. The optical depth halves on moving outward from one contour to the next. Baade's Window at $(\ell = 1.0°, b = -3.9°)$ is marked with a cross. Here, the optical depth has been measured to be $\sim 3.3 \pm 1.2 \times 10^{-6}$ by Ogle. Also marked with a cross is the window at $(\ell = 2.55°, b = -3.64°)$ where the optical depth to the red clump stars is measured to be $\sim 3.9 \times 10^{-6}$ by Macho. As can by seen by comparison with the contours, this axisymmetric model cannot reproduce such high values.

Paczyński[12] predicted the optical depth towards the Bulge was $\sim 10^{-6}$. It therefore came as a surprise when the experimentalists reported a much higher optical depth. For example, Alcock et al.[15] have already published a dataset of 45 candidate events from an analysis of 12.6 million stars over 190 days. They estimated the optical depth of the red clump stars in the Bulge region is $\tau \sim 3.9^{+1.8}_{-1.2} \times 10^{-6}$. This is an average over several square degrees centered on $(\ell = 2.55°, b = -3.64°)$. The optical depth of the full sample has the lower value of $\tau \sim 2.4 \pm .5 \times 10^{-6}$. The optical depth has also been measured at Baade's Window $(\ell = 1.0°, b = -3.9°)$ by the Ogle group[16] as $\tau \sim 3.3 \pm 1.2 \times 10^{-6}$. These high values pose grave difficulties for axisymmetric models of the inner Galaxy. This is illustrated in Fig. 6, which depicts a microlensing map[31] – that is, contours of equal optical depth in the plane of Galactic longitude and latitude. The optical depth halves on passing from one contour outward to the next. Of course, the Bulge is heavily obscured at visible wavelengths – two of the windows for which the microlensing searches have reported results are marked with crosses. The actual model shown in Fig. 6 is an axisymmetric Bulge and disk. The local circular speed has been pushed up to $\sim 240\,\mathrm{kms}^{-1}$, almost as high as is possible. The Bulge has been made as massive as is consistent with the rotation curve in the inner Galaxy. It is normalised to yield a circular velocity of $260\,\mathrm{kms}^{-1}$ at $500\,\mathrm{pc}$. The total mass of the bulge and spheroid integrated to infinity is $3.9 \times 10^{10}\,\mathrm{M}_\odot$. The Galactic disk is modelled by a spheroidally stratified exponential disk with a local column density of $80\,\mathrm{M}_\odot\,\mathrm{pc}^{-2}$, which is a higher than warranted by studies of the vertical kinematics of tracer stars, but perhaps still just possible. Its mass is $5.6 \times 10^{10}\,\mathrm{M}_\odot$. Despite pushing all constraints to the extremes of their possible ranges, this axisymmetric model can barely reproduce the high optical depths seen by the experimentalists, For example, the optical depth at Baade's Window is $2.1 \times 10^{-6}$, which is just consistent with the lower limit on the Ogle results.

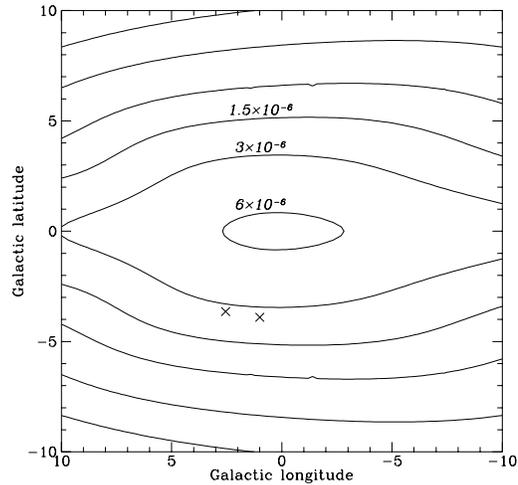

Figure 7: Contours of equal optical depth in the inner Galaxy, assuming a barred model with a canonical disk. The bar is Dwek et al.'s best fit to the COBE/DIRBE infrared surface photometry (the so-called G2 model). It is a triaxial Gaussian density distribution, the major axis of which is viewed at an angle of $13.4°$ degrees to the axis joining the Sun to the Galactic Center. The optical depth at Baade's Window is $2.52 \times 10^{-6}$.

Paczynski et al.[32] suggested that the high optical depth may be the signature of a bar, viewed somewhat pole-on. Part of the reasoning that led to this proposal was the failure of axisymmetric models to reproduce the microlensing data. Of course, the idea that the Galaxy is barred is quite old – the first suggestion goes back at least to de Vaucouleurs in 1964. The modern ideas of the barred structure of the inner Galaxy owe much to Binney, Gerhard, Spergel and co-workers[33,34,35]. The evidence is now impressive – for example, starcounts, kinematics of neutral and ionised gas, the COBE/DIRBE surface photometry all show the tell-tale evidence of a triaxial bar. Fig. 7 depicts a microlensing map for a barred model of the inner Galaxy. The bar is the best-fitting model to the COBE/DIRBE photometry found by Dwek et al.[36] The viewing angle is $\sim 13.4°$. This model certainly comes closer to reproducing the high optical depth to microlensing. The optical depth at $(\ell = 2.55°, b = -3.64°)$ is $\tau = 2.8 \times 10^{-6}$, whereas the optical depth at Baade's Window is $\tau = 2.52 \times 10^{-6}$. This is still lower than the data, but probably consistent given all the uncertainties. If the long axis of the bar is pointing nearly along the line-of-sight, then the optical depth to microlensing is enhanced in barred models over axisymmetric models.[31,37]

All investigators agree that an almost pole-on bar can augment the optical depth. But, it is unclear whether this is the correct interpretation of the data. The most reliable luminosity distribution for the bar is provided by Binney, Gerhard & Spergel.[34] These workers applied a Lucy-Richardson algorithm to the cleaned and corrected infra-red surface photometry to derive a three dimensional luminosity distribution. This extends Dwek et al.'s[36] analysis in two ways – first, the fitting method is non-parametric and second, the modelling of the extinction is more sophisticated than a simple screen. Bissantz et al.[38] analyse the asymmetries of the projections of these models and compare with the COBE/DIRBE data. They conclude that the Galactic bar is probably not seen pole-on and is anyway not highly elongated. The mass to light ratio of the model is found by comparison of the terminal velocities of SPH simulated gas flow with the observed HI and CO data. They find that the optical depth in Baade's Window lies in the range $0.83 \times 10^{-6} \lesssim \tau \lesssim 0.89 \times 10^{-6}$ for main sequence stars and in the range $1.2 \times 10^{-6} \lesssim \tau \lesssim 1.4 \times 10^{-6}$ for red clump stars. These results are much lower than the observations. The resolution of this seeming contradiction is not yet clear. One possibility is that the mass to light ratio varies sharply with height above the Galactic plane.[38] Another possibility is that the the microlensing experiments are overestimating the optical depth, as suggested on other grounds by Alard.[39]

To model the rate, a distribution of stellar orbits that builds the bar is needed. Unfortunately, stellar dynamical models of bars – rotating three di-mensional objects – are in their infancy. Zhao [40] deserves credit for being the first to apply the numerical method of Schwarzschild to the problem. His model gives a realistic bar distribution function that reproduces the optical depth, as well as the stellar kinematics at selected windows. But, there is still ample scope for more theoretical investment in this fundamental problem. The need for more and better barred representations of the inner Galaxy has led a group of us at Oxford (Häfner, Binney, Evans, Dehnen) to explore this problem more systematically by building sequences of models, whose stellar orbits add up to that of the bar (as inferred from the deprojected, cleaned and corrected surface photometry). There is hope that we will be able to make several very realistic barred models available in the near future.

## 5  Conclusions

Fairly definite conclusions can be made from the existing data-set on microlensing events towards the Large Magellanic Cloud. The halo cannot be primarily made from objects in the mass range $10^{-7} \, M_\odot - 0.1 \, M_\odot$. The microlensing experiments have assuredly detected a baryonic component to the halo. Let us call the fraction of baryonic matter $f$. Then, if the disk is canonical, $f \sim 0.4$; if the disk is maximal $f \sim 0.5$. The masses of the deflectors probably lie in the range $0.05 \, M_\odot - 0.1 \, M_\odot$. The low estimates are given by halo models with radially anisotropic velocity distributions.

The correct interpretation of the high optical depth to microlensing towards the Bulge remains unclear. Undoubtedly, the inner Galaxy is barred – but it seems premature to conclude that the bar has been detected in the microlensing sky. If the bar is elongated and viewed almost down its long axis, then it is the direct cause of the enhanced optical depth. But, the best estimates for the three-dimensional infrared luminosity density in the inner Galaxy do not support this idea. Possible resolutions of the contradiction are that the infrared light distribution does not trace the mass or that the optical depth reported by the microlensing experiments is being overestimated.